\documentstyle[preprint,aps]{revtex}

\begin{document}
\title
{Finite - temperature quantum phase transition in d - waves
superconductors}
\author{M. Crisan, D. Bodea, I. Grosu, and I. Tifrea\cite{iowa}}
\address{Department of Theoretical Physics, University of Cluj, 3400
Cluj-Napoca, Romania}
\maketitle

\begin{abstract}
The zero temperature d - wave superconductor phase transition
theory given in the case of $T=0$ for two - dimensional
superconductors (I. Herbut, PRL {\bf 85}, 1532 (2000)) is
generalized for finite temperatures. The Gaussian behavior of the
system is associated with a non - Fermi behavior of the normal
state observed in the resistivity of cuprate superconductors.
\end{abstract}

\newpage
\narrowtext

\section{Model and Scaling Equations}
In a fermionic system disorder and the attractive interaction on
the d - wave channel at $T=0$ give rise to the superconductor
pairing. The two - dimensional model ($d=2$) has been studied in
\cite{1} and the main result of this paper is the maping of the
fermionic system in a dissipative bosonic system with a wide
crossover regime controlled by the fluctuations of the order
parameter. This system is similar to the insulator -
superconductor at $T=0$ and finite temperature reconsidered by the
Urbana group \cite{2,3,4,5,6}, in study of the transport
properties near the critical point of this transition. Following
classical results \cite{7} from the superfluid - insulator
transition they showed that the insulator - superconductor
transition is characterized by the dynamical critical exponent
$z=2$ and used the renormalization - group method (RG) to describe
the critical behavior of the conductivity near the transition
point. The model used in \cite{1} contain an action with
dissipative term which is a relevant perturbation with the
particle - hole asymmetry. In fact, the analogy between the d -
wave superconductivity and this system was mentioned in connection
with the contribution of the different regimes to the
conductivity.

This report is complementary to papers \cite{1} and \cite{5}
considering the crossover effects as well as the case of the
finite value of the damping parameter from the dissipative term of
action.

We start with the action from Ref. \cite{1} defined using the
field operators $\phi_i(k)$ written as:
\begin{eqnarray}
S[\phi] & = & \sum_i \; \sum_k \; \left[ \phi_i^{\dagger}(k) \;
\left( \frac{\hbar^2 k^2}{2 m} + \frac{|\omega_n|}{\Gamma} - \mu
\right) \; \phi_i(k) \right] \nonumber \\
& + & \frac{u}{4} \; \sum_i \; \sum_{k_1} \ldots \sum_{k_4} \;
\delta \left( \sum_{l=1}^4 k_l \right) \; \phi_i(k_1) \ldots
\phi_i(k_4) \label{1}
\end{eqnarray}
In Eq. (\ref{1}) we have contributions of the n - independent
replicas and $\phi_i(k)=\phi(k)$. The other parameters are related
with those from Ref. \cite{1} as: $\mu = \mu_b m_b^{1/2}$, $m =
m+b^{1/2}$, $u = 4 \lambda$ and $\Gamma$ is an energy parameter
which controls the strength of the quantum fluctuations. The
effective chemical potential $\mu_b$ which is negative for the
normal phase ( see Ref. \cite{1}) is expressed by the relation
$\mu_b = (\ln \tau E_F - g^{-1})/\tau$, $\tau$ being the
scattering time, proportional to the impurities concentration and
$g$ the superconductor coupling constant. If we introduce the
critical scattering time $\tau_c = 2/ \Delta(0)$ ($\Delta(0)$ is
the $T=0$ order parameter) above which the order parameter
vanishes, the chemical potential $\mu_b$ depends of the impurities
concentration, $x$, as $\mu_b(x) \sim |x-x_c|$. This dependence is
obtained using the wellknown result that the scattering on a non -
magnetic impurities in d - wave superconductors has a similar
effect with the scattering on the magnetic impurities in the s -
wave superconductors, and also using $1/\tau \sim x$.

In this model the Eq. (\ref{1}) describe a quantum phase
transition controlled by non - magnetic impurities in d - wave
superconductors.

The effect of finite temperature on this quantum phase transition
can be studied using the RG method and has been applied by
different authors \cite{7,8,9,10,11,12,13,14,15} to study the
influence of temperature on the quantum phase transition in the
interacting bosonic systems. Following this method we consider the
case of finite temperatures and $\Gamma \ne 0$ and using the
scaling $k=k'/b$ and $\omega_n = \omega_n' / b^z$ ($b=\ln l$ and
$z$ is the critical exponent) we get the scaling equations:
\begin{equation}
\frac{d T(l)}{d l} \; = \; 2 T(l) \label{2}
\end{equation}
\begin{equation}
\frac{d \Gamma(l)}{d l} \; = \; 0 \label{3}
\end{equation}
\begin{equation}
\frac{d \mu(l)}{d l} \; = \; 2 \mu(l) \; - \; f^{(2)}
[T(l),\mu(l)] \; u(l) \label{4}
\end{equation}
\begin{equation}
\frac{d u(l)}{d l} \; = \; (2-z) u(l) \; - \; f^{(4)}[T(l),\mu(l)]
\; u^2(l) \label{5}
\end{equation}
and for the free energy
\begin{equation}
\frac{d F(l)}{d l} \; = \; 4 F(l) \; + \; f^{(0)}[T(l),\mu(l)]
\label{6}
\end{equation}
Functions $f^{(2)}$, $f^{(4)}$ and $f^{(0)}$ are given for $d=2$
system by:
\begin{equation}
f^{(2)}[T(l),\mu(l)] \; = \; \int_0^{\Lambda} \frac{d^2 {\bf
k}}{(2 \pi)^2} \; k_B T \; \sum_n \; \left[ \frac{\hbar^2 k^2}{2
m} + \frac{|\omega_n|}{\Gamma} - \mu \right]^{-1} \label{7}
\end{equation}
\begin{equation}
f^{(4)}[T(l),\mu(l)] \; = \; \int_0^{\Lambda} \frac{d^2 {\bf
k}}{(2 \pi)^2} \; k_B T \; \sum_n \; \left[ \frac{\hbar^2 k^2}{2
m} + \frac{|\omega_n|}{\Gamma} - \mu \right]^{-2} \label{8}
\end{equation}
\begin{equation}
f^{(0)}[T(l),\mu(l)] \; = \; \int_0^{\Lambda} \frac{d^2 {\bf
k}}{(2 \pi)^2} \; k_B T \; \sum_n \; \arctan \frac{Im \; \chi({\bf
k},i \omega_n)}{Re \; \chi({\bf k},i \omega_n)} \label{9}
\end{equation}
where
\begin{displaymath}
\chi^{-1}({\bf k},i \omega_n) \; = \; \frac{\hbar^2 k^2}{2 m} +
\frac{|\omega_n|}{\Gamma} - \mu
\end{displaymath}
and $\Lambda$ is the momentum cutoff.

We have to mention that these equations have been obtained from
Eq. (\ref{1}) neglecting the quartic term which describes the
interaction between fluctuations from different replicas.

\section{Fixed points}

In order to study the influence of the temperature on the quantum
effects, we have to calculate the fixed points of the scaling
equations.

As in the case of the bosons, (see Ref. \cite{12}) we will analyze
the two relevant cases: the low temperature and high temperature
regimes.

\subsection{Low temperature regime}

In the low temperature regime $z=2$ and from Eqs.
(\ref{2}-\ref{5}) we get:
\begin{equation}
\frac{d u(l)}{d l} \; = \; - \; \frac{m}{4 \pi \hbar^2} \; u^2(l)
\label{10}
\end{equation}
\begin{equation}
\frac{d \mu(l)}{d l} \; = \; 2 \mu(l) \; + \; f^{(2)}
[T(l),\mu(l)] \; u(l) \label{11}
\end{equation}
where we used $\mu <0$ for the normal phase.

The function $f^{(2)}$ has been calculated taking the $\omega_n
=0$ and $\omega_n \ne 0$ frequencies at finite temperatures as:
\begin{equation}
f^{(2)}[T(l),\mu(l)] \; \simeq \; \frac{\Lambda^2}{4 \pi} \; + \;
\frac{m k_B T}{2 \pi \hbar^2} \; \int_0^{\infty} \; \frac{d
y}{\exp \left[ y + \frac{\mu}{T} \right] - 1} \label{12}
\end{equation}
From Eq. (\ref{10}) we obtain:
\begin{equation}
u(l) \; = \; \frac{4 \pi \hbar^2}{m} \; \frac{1}{l + l_0}
\label{13}
\end{equation}
where $l_0 = 4 \pi \hbar^2 / m u$.

In order to get the fixed point of Eq. (\ref{11}) we use the
condition $d \mu(l)/dl = 0$ and we obtain:
\begin{equation}
2 \mu_0^* \; + \; \frac{\Lambda^2 u}{4 \pi} \; + \; \frac{m k_B T
}{2 \pi \hbar^2} \; \int_0^{\infty} \frac{d y}{\exp \left[ y +
\frac{\mu}{T} \right] -1} \; u(l) \; = \; 0 \label{14}
\end{equation}
From this equation we can see that for finite temperature there is
no phase transition because the last integral from Eq. (\ref{14})
is divergent. At $T=0$ we get:
\begin{equation}
\mu_0^* \; = \; - \; \frac{\Lambda^2 u}{8 \pi} \label{15}
\end{equation}
which correspond to the quantum phase transition.

\subsection{High temperatures regime}

In this case the scaling equations are obtained from general
equations (\ref{2}-\ref{5}) by taking $z=2$ and using the
approximation $\coth x/2T \simeq 2T/x$. The scaling equations (see
also Ref. \cite{13,14}) becomes:
\begin{equation}
\frac{d u(l)}{d l} \; = \; 2 u(l) \; - \; \frac{5}{2} K_2
\frac{\Lambda^2 k_B T(l)}{[\epsilon_{\Lambda} + \mu]^2} \label{16}
\end{equation}
\begin{equation}
\frac{d \mu(l)}{d l} \; = \; 2 \mu(l) \; - \; \frac{1}{2} K_2
\frac{\Lambda^2 k_B T(l)}{[\epsilon_{\Lambda} + \mu]^2} \label{17}
\end{equation}
where $\epsilon_{\Lambda} = \hbar^2 \Lambda^2 /2 m$ and $K_2 = 1/2
\pi $. From these equations we obtain the fixed points $\mu^* = -
5 \epsilon_{\Lambda}/2$ and $u^* = 2 \pi \epsilon_{\Lambda}^2 / 45
\Lambda^2 k_B T$. This is the $T = \infty$ fixed point.

The crossover between the two regimes is equivalent to the
crossover between the quantum and classical critical behavior and
we will define a parameter $l^*$ for the two regimes.

\section{Crossover between quantum and classical regimes}

The quantum behavior is defined by the fixed point $T=0$. The
effect of the temperature can be considered if we determine the
temperature scale $\tilde{l}$ defined by:
\begin{equation}
T(\tilde{l}) \; = \; T_0 \label{18}
\end{equation}
where $T_0$ is a characteristic temperature. Using the solution
$T(l) = T e^{2l}$ we get the parameter $\tilde{l}$ as:
\begin{equation}
\tilde{l} \; = \; \frac{1}{2} \ln \frac{T_0}{T} \label{19}
\end{equation}
Using Eqs. (\ref{10}-\ref{11}) we define the parameter $l_0^*$ by:
\begin{equation}
\mu(l_0^*) \; = \; - \alpha \frac{\hbar^2 \Lambda^2}{2 m}
\label{20}
\end{equation}
being the energy parameter which define the low temperature
critical region, and $\alpha \le 1$.

From Eq. (\ref{11}) we calculate (see also Ref. \cite{12})
$\mu(l)$ as:
\begin{equation}
\mu(l) \; = \; - \frac{4 \hbar^2 \Lambda^2}{2 m} \; e^{2l}
\int_0^l \frac{d l'}{l'+l_0} \; \frac{e^{-2l'}}{\exp \left[
\frac{\hbar^2 \Lambda^2}{2 m k_B T} e^{-2l'} \right] -1}
\label{21}
\end{equation}
This expression can be transformed in:
\begin{eqnarray}
\mu(l) & = & - 4 \epsilon_{\Lambda} e^{2l} \; \left\{ \frac{k_B
T}{\epsilon_{\Lambda}} \left[ \frac{1}{2 l_0} \ln \left(1 - e^{-
\frac{\epsilon_{\Lambda}}{k_B T}} \right) \right. \right.
\nonumber \\
& - & \left. \left. \frac{1}{2 l_0} \left(1 + \frac{l}{l_0}
\right)^{-1} \; \ln \left( 1 - e^{- \frac{\epsilon_{\Lambda}}{k_B
T }e^{-2l}} \right) \right] \; - \; \frac{k_B
T}{\epsilon_{\Lambda}} F(l) \right\} \label{22}
\end{eqnarray}
where
\begin{equation}
F(l)\; = \; \int_0^{2l} \frac{d x}{(x+2l_0)^2} \ln \left[ 1 - \exp
\left( - \frac{\epsilon_{\Lambda}}{k_B T} e^{-2l} \right) \right]
\label{23}
\end{equation}
The energy scale will be fixed using the Eqs. (\ref{12}) and
(\ref{21}) and following Ref. \cite{} we calculate
\begin{equation}
e^{-2l^*} \; \simeq \; 16 \frac{T}{T_0} \; \frac{1}{\ln
\frac{T}{T_0}} \label{24}
\end{equation}
where $T_0 = \alpha \hbar^2 \Lambda^2 / 8 m k_B$. Using this
result we obtain
\begin{equation}
l_0^* \; = \; \frac{1}{2} \; \ln \left( \frac{T_0}{T} \ln
\frac{T_0}{T} \right) \label{25}
\end{equation}

Let us consider the high temperatures regime, describe by the Eqs.
(\ref{16}-\ref{17}), which will be rewritten using the new
interaction $v(l) = k_B T(l) u(l)$, as
\begin{equation}
\frac{d v(l)}{d l} \; = \; 2 v(l) \; - \; \frac{5}{\pi}
\frac{m^2}{\hbar^4 \Lambda^2} \; v^2(l) \label{26}
\end{equation}
\begin{equation}
\frac{d \mu(l)}{d l} \; = \; 2 \mu(l) \; - \; \frac{m}{\pi \hbar^2
\Lambda^2} \; v(l) \label{27}
\end{equation}
The exact solution of Eq. (\ref{26}),
\begin{equation}
v(l) \; = \; \frac{2 v(\tilde{l})}{B v(\tilde{l}) + [2 -
v(\tilde{l})] e^{-2(l-\tilde{l})}} \label{28}
\end{equation}
where $B = 5 m^2 / \pi \hbar^4 \Lambda^2$ will be approximated as
\begin{equation}
v(l) \; \simeq \; v(\tilde{l}) \; e^{-2(l-\tilde{l})} \label{29}
\end{equation}
and from Eq. (\ref{27}) we get
\begin{equation}
\mu(l) \; = \; - \frac{4 (l-\tilde{l})}{\tilde{l}+l_0} \;
e^{2(l-\tilde{l})} \label{30}
\end{equation}
The new energy scale will be fixed by $l_1^*$ as
\begin{equation}
\mu(l_1^*) \; = \; - \alpha \frac{\hbar^2 \Lambda^2}{2 m}
\label{31}
\end{equation}
and from Eqs. (\ref{30}) and (\ref{31}) we express
\begin{equation}
l_1^* \; = \; \frac{1}{2} \; \ln \left( \frac{T_0}{T} \ln
\frac{T_0}{T} \right) \label{32}
\end{equation}
The Eqs. (\ref{25}) and (\ref{32}) show that we can perform a
matching between the two regimes and:
\begin{equation}
l_0^* \; = \; l_1^* \; = \; l^* \label{33}
\end{equation}
Physically this result can be regarded as following: In the $T$ -
$l$ plane we can reach the critical region starting from the
quantum regime or from the classical regime. The characteristic
temperature in this point is
\begin{equation}
T^* \; = \; T_0 \; \ln \frac{T_0}{T} \label{34}
\end{equation}
and $u^*(l=l^*)$ calculated from Eq. (\ref{13}) is
\begin{equation}
u^* \; = \; \frac{4 \pi \hbar^2}{m} \; \frac{1}{l_0 + \ln \left(
\frac{T_0}{T} \ln \frac{T_0}{T} \right)} \label{35}
\end{equation}
This value is very small for $T<T_0$ and in this region the
perturbation theory is valid.

In the next section we calculate the specific heat in these two
regimes.

\section{Free energy and specific heat}

The free energy will be calculated from Eqs. (\ref{6}) and
(\ref{9}) as:
\begin{equation}
\frac{d F}{d l} \; = \; 4 F(l) \; + \; f^{(0)}[T(l),\mu(l)]
\label{36}
\end{equation}
where
\begin{equation}
f^{(0)}[T(l),\mu(l)] \; = \; - \frac{K_2 \Lambda^2}{\pi} \;
\int_0^{\Gamma} d \omega \; \coth \frac{\omega}{2 k_B T(l)} \;
\tan^{-1} \frac{\omega}{\left[ \frac{\hbar^2 \Lambda^2}{2 m} -
\mu(l) \right]} \label{37}
\end{equation}
In order to get the temperature dependence of the free energy and
the specific heat we divide the temperature interval in two
regimes: $0 \le x \le \frac{1}{2} \ln \frac{T_0}{T}$ (quantum
regime) and $\frac{1}{2} \ln \frac{T_0}{T} \le x \le l_M$, where
$l_M$ is temperature independent in the classical regime. The
general solution of Eq. (\ref{36}) has the form:
\begin{equation}
F[T(l)] \; = \; \int_0^l dx \; e^{-4x} \; f^{(0)}[T e^{2x}]
\label{38}
\end{equation}
and following \cite{10} we expand $f^{(0)}(T)$ as $\lim_{T
\rightarrow 0} [f^{(0)}(T) - f^{(0)}(0)] \sim T^2$. In the first
regime the contribution of $f^{(0)}(0)$ is negligible. In the
second regime one may approximate $f^{(0)}(T) \sim T$ and from Eq.
(\ref{28}) we get
\begin{equation}
F(T) \; = \; F_1 T^2 \; \ln \frac{T_0}{T} \; - \; F_2 T \label{39}
\end{equation}
where $F_1$ and $F_2$ are constants.

Using now for the specific heat the relation $C_v = - k_B T
\partial^2 F/ \partial T^2$ we obtain
\begin{equation}
C_v(T) \; = \; \gamma_0 T \; + \; \gamma_1 T \ln \frac{T}{T_0}
\label{40}
\end{equation}
where the first term gives the classic contribution and the last
term is the contribution of the non - Fermi excitations.

\section{Discussions}

The results obtained can be discussed refereing to the
experimental result obtained on the $Zn$ substitution in cuprate
superconductors. The measurements \cite{16} predicted for zero
resistivity a value close to the universal $2D$ $\rho_0 \simeq
h/4e^2$ and a superconductor - insulator transition. More recent
experiments \cite{17} showed that the suppression of $d$ - wave
superconductivity leads to a metallic non - superconductor phase
and the metal - insulator transition is suggested at $k_F l \sim
1$. At low temperatures $\rho_{ab} \sim 1/T$ and $\rho(0)$ is
finite.

The temperature dependence of the specific heat is also different
from the behavior of the natural phase \cite{18}, but this was
explained by an energy dependent electronic density of states.
However, in the very low temperature domain, $T < 1 K$ this
dependence can be described by the quantum contribution $T \ln
T/T_0$ from Eq. (\ref{40}).

The possibility of a non - Fermi metallic state has been also
predicted recently \cite{19} for a $2D$ system with a field -
tuned superconductor - insulator transition. Just above the
transition, the phase appear to be metallic, but with strong
deviation from the Fermi - liquid behavior. Recently \cite{20},
the thermal conductivity measurements in these systems showed a
metal - insulator crossover for the normal state and the existence
of a low - energy scale in these materials.

The transport properties for such a model, but taking the
dissipative term zero, have been performed using RG in \cite{5}.
The difference between our results and the results from Ref.
\cite{5} are given by the approximation in the calculation of
$l^*$. The normal state has been considered in \cite{1} as a metal
state and the conductivity of this system behaves like in the two
- dimensional metal insulator transition. The dissipation term has
in this case the leading role and the singular part of the
conductivity calculated in \cite{21,22,23} is not universal just
for the case of the particle - hole symmetry. However, the
conductivity can be interpolated between this regime and the non -
universal case observed by different authors \cite{17,18}.

An interesting physical picture of the transport was developed by
Dalidovich and Philips \cite{2,3,4,5,6} using the standard theory
\cite{21,22} in the critical region. Such a theory is very close
to our picture excepting the technical aspects. However, we have
to mention that there is an important difference between our RG
calculation of $l^*$ and that from Ref. \cite{5}. It is given by
the fact that in Ref. \cite{5} this quantity has been obtained
using the results from \cite{15} where all the calculations have
been performed with a singular coupling constant. We showed in
\cite{12} that the calculation of $l^*$ has to be done using
$u(l)=1/(l+l_0)$, a value which gives a correct critical value for
the superfluid temperature. The effect of this difference on the
values of the physical observable as conductivity, will be
evaluated in a future paper, but preliminary numerical calculation
showed that in the low temperature regime the difference is very
small.

\end{document}